\documentclass[preprint,showpacs,preprintnumbers,amsmath,amssymb]{revtex4}

\usepackage{graphicx}
\usepackage{dcolumn}
\usepackage{bm}
\usepackage{amsmath}

\begin{document}

\preprint{}

\title{Dynamics of Quantum Dot Photonic Crystal Lasers}

\author{Bryan Ellis}
\email{bryane@stanford.edu}
\author{Ilya Fushman}
\author{Dirk Englund} 
\author{Bingyang Zhang}
\author{Yoshihisa Yamamoto}
\author{Jelena Vu\v{c}kovi\'{c}} 
 \affiliation{E. L. Ginzton Laboratory, Stanford University, Stanford, CA, 94305}

\date{\today}

\begin{abstract}

Quantum dot photonic crystal membrane lasers were fabricated and the large signal modulation characteristics were studied.  We find that the modulation characteristics of quantum dot lasers can be significantly improved using cavities with large spontaneous emission coupling factor.  Our experiments show, and simulations confirm, that the modulation rate is limited by the rate of carrier capture into the dots to around 30GHz in our present system.  
\end{abstract}

\maketitle

Recent advances in microfabrication technology have enabled researchers to control the electromagnetic environment and, consequently, the spontaneous emission rate of an emitter by coupling it to a semiconductor microcavity \cite{ref:Gerard_micropillar,ref:Englund_enhancement}.  In a microcavity where the spontaneous emission rate is significantly enhanced, a large fraction of spontaneously emitted photons can be coupled into a single optical mode.  This unique property has been used to demonstrate microcavity semiconductor lasers with thresholds dramatically reduced relative to conventional lasers \cite{ref:Loncar_PClaser}.  Thresholds can be further reduced by using quantum dots as the gain medium because of reduced active area and nonradiative recombination.  Recently, quantum dot lasers exhibiting high spontaneous emission coupling factors and very few quantum dots as a gain medium were demonstrated \cite{ref:Strauf_PClaser}.  In addition, microcavity lasers can be designed to have very broad modulation bandwidth because the relaxation oscillation can be shifted beyond the cavity cutoff frequency \cite{ref:Yamamoto_microcavity}.  A recent experiment demonstrated direct modulation rates far exceeding 100GHz in quantum well photonic crystal (PC) lasers \cite{ref:Altug_modulation}.  Lasers for applications such as high-speed optical telecommunications or on-chip optical interconnects could benefit from a combination of reduced thresholds and increased modulation bandwidth.  In this letter we examine the factors limiting the modulation rate of quantum dot photonic crystal lasers and demonstrate large-signal direct modulation rates approaching 30GHz in quantum dot structures.    

In a quantum dot laser the maximum modulation bandwidth is limited by either the frequency of the relaxation oscillations or the rate of carrier capture into the quantum dots depending on which is smaller.  In conventional quantum dot lasers at low pump powers, the relaxation oscillation frequency is significantly smaller than the rate of carrier capture into the dots.  This frequency increases with input power, so the modulation bandwidth can be enhanced by increasing pump power.  This technique was used to demonstrate small-signal modulation rates of several tens of GHz \cite{ref:Fathpour_tunnel}, but relatively large pump powers were necessary making these lasers impractical for low power applications.  In addition the large signal modulation rates are considerably slower because the turn-on delay times are on the order of a nanosecond \cite{ref:Grundmann_turnon,ref:Bhattacharya_turnon}.  This could be resolved by using a microcavity laser with enhanced spontaneous emission coupling factor, which increases the relaxation oscillation frequency (as demonstrated for quantum well lasers \cite{ref:Altug_modulation}).  In that case, the maximum modulation rate would be limited by the rate of carrier capture into the dots at practically achievable pump powers.  To demonstrate the utility of this approach we fabricated quantum dot photonic crystal lasers and investigated their large-signal modulation characteristics.  We find that the maximum modulation rate is limited by the rate of carrier capture into the dots, as predicted.

\begin{figure}[h]
		\includegraphics{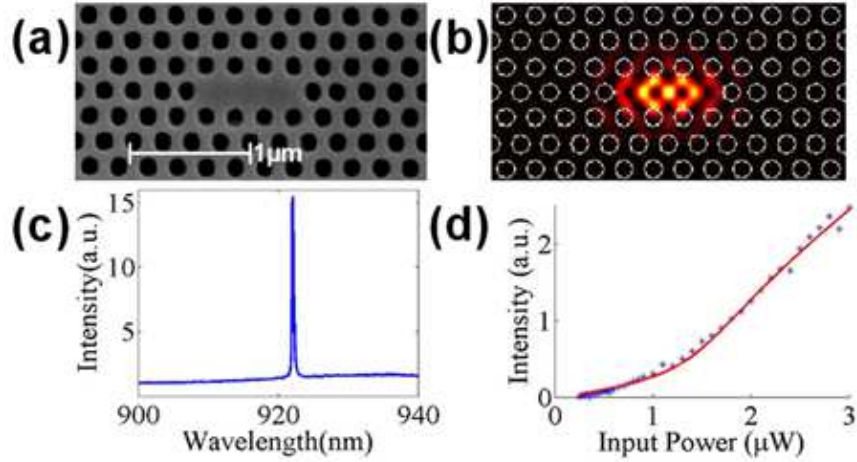}
		\caption{(Color online) \textbf{(a):} Scanning electron microscope image of fabricated laser cavity \textbf{(b):} Finite-difference time domain simulation of electric field amplitude for high-Q cavity mode \textbf{(c):} Spectrum of laser above threshold.  \textbf{ (d):} Light in-Light out curve of one PC laser (blue) and fit to rate equations (red) demonstrating $\beta$ of approximately 0.2}
\end{figure}

The laser cavities employed in our experiment are high-Q linear 3-hole defect PC cavities in a GaAs membrane (Fig. 1a).  Finite-Difference Time Domain simulations were used to design the cavities to have a high Q resonance near the center of the quantum dot gain spectrum (Fig. 1b).  The membrane is approximately 135nm thick and contains one layer of high density (600/$\mu m^{2}$) InAs quantum dots.  Electron beam lithography is used to define the photonic crystal pattern in PMMA.  A Chlorine-based reactive ion etch is used to create the holes.  Finally an Al$_{0.9}$Ga$_{0.1}$As sacrificial layer is first oxidized, then removed in a KOH solution to release the membrane.  After fabrication the structures are placed inside a He-flow cryostat and cooled to 5K (necessary for operation of the InAs/GaAs quantum dots with shallow quantum confinement).  The lasers are optically pumped using a Ti-Sapphire laser, and the emission is detected using a liquid nitrogen cooled spectrometer or streak camera.

To determine the cavity photon lifetime $\tau_p$, the quality factor of the cavities was measured well below threshold using continuous wave pumping.  Fits to a Lorentzian lineshape indicate that the cold-cavity quality factors are around 3000, corresponding to a cavity photon lifetime of about 1.5ps.  To investigate the dynamics of the structures, we pumped them with 3ps pulses at an 80MHz repetition rate using the Ti-Sapphire laser.  Streak camera measurements of the rise time of photoluminescence from quantum dots in bulk GaAs indicate that the carrier capture time is around 10ps for a wide range of pump powers.  Because the carrier capture time is longer than the cavity photon lifetime, it will ultimately determine the maximum modulation bandwidth of the lasers.  Figure 1c shows an L-L curve taken under pulsed pumping conditions and a fit to the rate equations for one of the PC lasers.   The L-L curve exhibits a threshold of around 1$\mu W$ confirming that the structures are lasing.  In the best structures (where the cavity mode is near the center of the gain spectrum) threshold values are around 250nW average power, while in other structures with more absorption and less gain, threshold values were measured at several $\mu W$ average power.  From fits to the light in-light out curve we estimate that in our structures the spontaneous emission coupling factor $\beta$ is around 0.2.  To confirm that the spontaneous emission rate in our cavities is significantly enhanced we used a streak camera to compare the decay time of photoluminescence from quantum dots in bulk GaAs and cavity coupled dots in nonlasing devices.  The measurements show that the dot lifetime is significantly reduced from the bulk value of 2.5ns to 300ps when coupled to the PC cavity.

To investigate the large-signal modulation response, emission from the lasers above threshold was collected by the streak camera.  One of the main advantages of cavity-QED enhanced lasers is the decreased rise time because spontaneous emission rapidly builds up the photon number in the laser mode \cite{ref:Altug_modulation}.  Experimentally we find that as pump power is increased the rise time is reduced to 13.5ps when the laser is pumped at about 5 times threshold (Figure 2b).  Experiments performed at 10 and 15 times threshold indicate that the rise time is pinned at about 12ps even at very high pump powers.  From simulations we conclude that the rise time of quantum dot lasers is limited by the carrier capture time.  However, in high-$\beta$ lasers this limit is practically achievable because it is approached at lower pump powers relative to threshold (as opposed to quantum dot lasers not employing stong cavity effects where higher power pumping is needed).

\begin{figure}[!h]
	\includegraphics{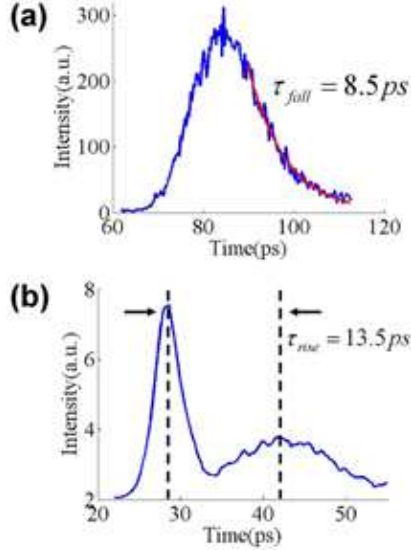}
  \caption{(Color online) \textbf{(a)} Laser response pumped at around 5 times threshold (blue) and exponential fit (red) demonstrating a fall time of approximately 8.5ps. \textbf{(b)} Streak camera response showing the time delay between pump (first peak) and PC laser (second peak) reponse demonstrating a rise time of only 13.5ps.}
\end{figure}

Above threshold, higher pump powers lead to faster decay times due to increased stimulated emission rates.  Small-mode volume PC cavities can be used to achieve large photon densities and speed up this process.  Figure 3a shows the laser response at various pump powers, demonstrating the reduction in decay time with increasing pump power.  We observed a minimum decay time of 8.5ps at pump powers around 5 times threshold (Figure 2a).  For higher pump powers the laser response appears largely unchanged.  We attribute this to large carrier densities causing the gain to saturate preventing further decrease of the decay time, but more work is necessary to characterize saturation effects in our quantum dots.  Based on the measured laser response and the results of simulations, we predict our laser can be modulated at large-signal modulation rates approaching 30GHz.

\begin{figure}[!h]
	\includegraphics{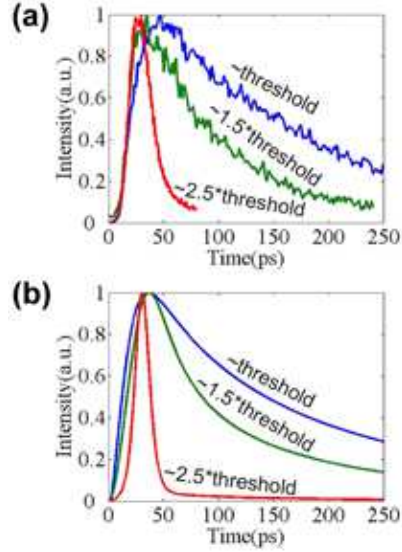}
  \caption{(Color online) \textbf{(a)} Experimentally measured laser response near threshold demonstrating reduction in fall time as the stimulated emission rate is increased \textbf{(b)} Simulated laser response for the same pumping conditions.
  }
\end{figure}

To accurately model the photonic crystal laser modulation characteristics, the usual rate equation model must be adapted to include the finite relaxation time from the wetting layer into the quantum dots.  We employed a three-level rate equation model adapted from \cite{ref:Coldren_Corzine,ref:OBrien_rateeq} for the photon density P, the quantum dot ground state carrier density N$_g$ and the wetting layer carrier density N$_w$:
\begin{equation}
\frac{dN_w}{dt}=R_p-\frac{N_w}{\tau_w}-\frac{N_w}{\tau_c}
\end{equation}
\begin{equation}
\frac{dN_g}{dt}=\frac{N_w}{\tau_c}-\frac{N_g}{\tau_{sp}}-\frac{N_g}{\tau_{nr}}-GP
\end{equation}
\begin{equation}
\frac{dP}{dt}=\Gamma G P+\Gamma \beta \frac{N_g}{\tau_{sp}}-\frac{P}{\tau_p}
\end{equation}
Here $\tau_{sp}$ and $\tau_{nr}$ are the spontaneous emission lifetime and nonradiative lifetime of the dots, $\tau_w$ is the lifetime of carriers in the wetting layer including spontaneous and nonradiative recombination, $\tau_p$ the photon lifetime, $\tau_c$ the carrier capture time into the dots, $\Gamma$ the confinement factor, and $R_p$ the pump rate.  Streak camera measurements of the wetting layer response indicate that for our samples $1/\tau_w \approx 1/100ps$.  From estimates of the overlap of the mode volume with the gain medium we estimate that $\Gamma$=0.028 in our lasers.  We have assumed a linear gain model
\begin{equation}
G=G_o*(N_g-N_{tr})
\end{equation}
where $G_o$ is the linear gain coefficient and $N_{tr}$ is the transparency carrier density.  From fits to the L-L curve we estimate that in our system $G_o = 8.13*10^{-6} s^{-1}$ and $N_{tr}=3.22*10^{17} cm^{-3}$.  The linear gain model was chosen to give good quantitative agreement when the lasers are operated around threshold, but the model will overestimate the gain well above threshold.  A more sophisticated gain model is necessary to model the modulation response of the lasers when gain saturation effects become significant.  Figure 4b shows the simulated laser response at various pump powers demonstrating good agreement between theory and experiment.

For practical applications, room temperature operation of the quantum dot microcavity lasers will be necessary.  Quantum dot lasers have been shown to have a strongly temperature-dependent modulation response.  Recent work on tunnel injection quantum dots has demonstrated that fast relaxation rates ($\sim 1.7ps$) at room temperature are achievable.  These dots have been used to demonstrate 25GHz small signal modulation bandwidth \cite{ref:Fathpour_tunnel}.  We believe this bandwidth can be significantly improved using a PC laser cavity.

In summary, we have investigated the large-signal modulation characteristics of quantum dot PC lasers.  We demonstrated that cavity-QED effects can be used to combine ultra-low threshold operation and improved bandwidth.  Because of their low-power consumption and high-speed operation, quantum dot microcavity structures have the potential to significantly improve optical interconnect and photonic integrated circuit technology.

Financial support for this work was provided by the MARCO Interconnect Focus Center, NSF grants No. ECS-0424080 and No. ECS-0421483, the Stanford Graduate Fellowship, and the NDSEG fellowship.  Financial assistance for B.Y. Zhang was provided in part by JST/SORST.



\begin{thebibliography}{widest-label}
	\bibitem{ref:Gerard_micropillar}
	J. G\'{e}rard, B. Semarge, B. Gayral, B. Legrand, E. Costard, and V. Thierry-Mieg, Phys. Rev.
Lett. \textbf{81}, 1110 (1998).

	\bibitem{ref:Englund_enhancement}
	D. Englund, D. Fattal, E. Waks, G. Solomon, B. Zhang, T. Nakaoka, Y. Arakawa, Y. Yamamoto,
and J. Vuckovic, Phys. Rev. Lett. \textbf{95}, 013904 (2005).

	\bibitem{ref:Loncar_PClaser}
	M. Loncar, T. Yoshie, A. Scherer, P. Gogna, and Y. Qiu, Appl. Phys. Lett. \textbf{81}, 2680 (2002).
	
	\bibitem{ref:Strauf_PClaser}
	S. Strauf, K. Hennessy, M. Rakher, Y. Choi, A. Badolato, L. Andreani, E. Hu, P. Petroff, and
D. Bouwmeester, Phys. Rev. Lett. \textbf{96}, 127404 (2006).

	\bibitem{ref:Yamamoto_microcavity}
	Y. Yamamoto, S. Machida, and G. Bjork, Phys. Rev. A \textbf{44}, 657 (1991).
	
	\bibitem{ref:Altug_modulation}
	H. Altug, D. Englund, and J. Vuckovic, Nature Phys. \textbf{2}, 484 (2006).
	
	\bibitem{ref:Fathpour_tunnel}
	S. Fathpour, Z. Mi, and P. Bhattacharya, J. Phys. D \textbf{38}, 2103 (2005).
	
	\bibitem{ref:Grundmann_turnon}
	M. Grundmann, Appl. Phys. Lett. \textbf{77}, 1428 (2000).
	
	\bibitem{ref:Bhattacharya_turnon}
	P. Bhattacharya, S. Ghosh, S. Pradhan, J. Singh, Z. Wu, J. Urayama, K. Kim, and T. Norris,
IEEE J. Quantum Electron. \textbf{39}, 952 (2003).

	\bibitem{ref:Coldren_Corzine}
	L. Coldren and S. Corzine, Diode Lasers and Photonic Integrated Circuits (John Wiley and
Sons, Inc., 1995).

	\bibitem{ref:OBrien_rateeq}
	D. O'Brien, S. Hegarty, G. Huyet, and A. Uskov, Opt. Lett. \textbf{29}, 1072 (2004).
\end{thebibliography}

\newpage	

\end{document}